# FORMATION OF BOSONIC COMPACT OBJECTS*

*EDWARD SEIDEL*
National Center for Supercomputing Applications
605 E. Springfield Ave., Champaign, Illinois 61820

and

*WAI-MO SUEN*
McDonnell Center for the Space Sciences, Department of Physics
Washington University, St. Louis, Missouri 63130

ABSTRACT

We showed that compact bosonic objects can be formed through a process we called gravitational cooling. A central issue in the subject of boson star is whether a classical field configuration, *e.g.*, one described by the Klein-Gordon equation, can collapse to form a compact star-like object, as there is apparently no dissipation in the Klein-Gordon equation. We demonstrated that there IS an efficient cooling mechanism to get rid of the kinetic energy for the formation of a compact object purely through the gravitational coupling, a mechanism universal to all self-graviting fields.

Implications of this mechanism are discussed, including the abundance of bosonic stars in the universe, and the possibility of ruling out the axion as a dark matter condidate.

## 1. Introduction

It has long been known that there exist compact self-gravitating soliton-like equilibrium configurations of bosonic fields.[1] The recent surge of interest in these solitonic objects is largely due to the suggestion that the dark matter could be bosonic in nature. Many particle theories predict that weakly interacting bosons are abundant in the universe,[2] and may have played a significant role in the evolution of the universe. The simplest example of a compact bosonic star is made up of a complex massive Klein-Gordon scalar field, with no self-interaction except through gravity:

$$\Box \phi + m^2 \phi = 0, \quad G_{\mu\nu} = 8\pi T_{\mu\nu}(\phi) \tag{1}$$

where $T_{\mu\nu}$ is the usual stress energy given by $\phi$ and its derivatives. The non-linear coupling through gravity makes possible the non-topological solitonic solutions.

However, the existence of a solitonic solution is not a guarantee that such an object can actually be formed in the universe. It is well-known that for self-gravitating fields described by Eq. (1), there is a gravitational instability analogous to the Jeans' instability. However, if there is no efficient cooling mechanism to get rid of the excess kinetic energy, as is apparently the case for a system described by Eq. (1), a collapse initiated by this instability generally only leads to a diffuse viralized "cloud", but

---

not compact object. This is an outstanding fundamental question in the study of the bosonic compact objects.[1]

We found that there is a dissipationless cooling mechanism which very efficiently leads to the formation of compact bosonic objects. This mechanism, which we call gravitational cooling, is similar to the violent relaxation of collisionless stellar systems.[3] Quite independent of the initial conditions, a scalar field configuration described by Eq. (1), will collapse to form a compact soliton star by ejecting part of the scalar field, carrying out the excess kinetic energy, through the non-linear self-gravitational interactions during the collapse. The cooling mechanism has been numerically analyzed in Ref. 4.

## 2. Implications

The immediate implication of the existence of the dissipationless gravitational cooling mechanism is that bosonic solitonic stars can be formed in our universe. If the dark matter is described by a classical bosonic field, it could be made up of such stars, provided that the localized clouds separated from the Hubble flow are not too massive (other wise a black hole may form). These clouds will begin to collapse under their own self gravity, at first in free fall. Then, as non-linear gravitational effects becomes important at higher densities, the gravitational cooling process becomes important, allowing the configuration to settle into a compact bosonic object.

The abundance of such solitonic stars with astrophysical mass but microscopic size has interesting consequences on galaxy formation, microwave background power spectrum, galaxy dynamics, formation of the first stars, among other things. A particularly intriguing possibility is that such a mechanism may rule out the axion as a dark matter candidate. The axion miniclusters of Hogan and Rees[5] have a density of $10^7$ g/cm$^3$. At such a density, the annihilation ($AA \to \gamma\gamma$), and other dissipative processes, are not effective,[6] and the evolution is accurately described by Eq. (1). Therefore the mini-clusters must continue on collapsing through gravitational cooling after separating out from the Hubble flow. For axions with mass $m \sim 10^{-5}$ eV, the maximum oscillaton mass is $10^{28}$ g, which is bigger than the total mass of a minicluster. Hence one might expect that the end point of the collapse is an oscillaton (but not a black hole). If there is no further fragmentation during the collapse of the minicluster, the resulting oscillaton has a density of $\rho = 10^{24}$ g/cm$^3$ (phase space number density $n_p \simeq 10^{62}$ cm$^{-3}$). However, at such high density, the axion is no longer described by a free Klein-Gordon field. In particular, we expect the stimulated decay of axions to be important (single axion decay rate is extremely small, $r \sim 10^{-49}$ $sec^{-1}$, for $m_a \sim 10^{-5}$ eV). The amplification coming from stimulated decay gives a factor of $\exp(D)$, with[7]

$$D \sim \frac{\Gamma_\pi M_p^2 V_e}{R m_\pi^4} \frac{f_\pi}{f_a} \qquad (2)$$

where $\Gamma_\pi \sim 8$ eV, $f_\pi \sim 1$ GeV, $f_a \sim 10^{12}$ GeV (for $m_a \sim 10^{-5}$ eV), $m_\pi = 135$ MeV, $V_e$ is the escape velocity ($\sim \sqrt{\frac{2GM}{R}}$), and $R$ the radius of the bound object. For an oscillaton with the mass of a mini-cluster, $D \sim 10^7$, which simply implies that the collapse driven by the gravitational cooling ends up in a bright flash.[8] This suggests that it might not be self-consistent to have the axions as a dark matter candidate: the axion has a tendency to form compact objects (oscillatons) in a short dynamical time scale, but it is unstable in such a state.

## Acknowledgements

We would like to thank Ram Cowsik, Leonid Grishchuk, Rocky Kolb, Martin Rees, Matt Vissor and Clifford Will for useful discussions. The work is supported by The National Center for Supercomputing Applications and by NSF grant Nos. 91-16682, 94-04788, and 94-07882.